# APLICACIÓN DE MODELOS DE LENGUAJE GPT PARA LA INNOVACIÓN EN ACTIVIDADES EN LA DOCENCIA UNIVERSITARIA*

# APPLICATION OF GPT LANGUAGE MODELS FOR INNOVATION IN ACTIVITIES IN UNIVERSITY TEACHING

MANUEL DE BUENAGA RODRÍGUEZ

FRANCISCO JAVIER BUENO GUILLÉN

*Departamento de Ciencias de Computación*

*Universidad de Alcalá*

**Resumen:**

Los modelos de lenguaje GPT (Generative Pre-trained Transformer) son una tecnología basada en inteligencia artificial y procesamiento del lenguaje natural que permite generar texto automáticamente. Existe un interés creciente en torno a la aplicación de modelos de lenguaje GPT a la docencia universitaria en diferentes dimensiones. Desde el punto de vista de la innovación en las actividades de los alumnos y profesores, pueden ofrecer soporte en la comprensión y generación de contenidos, la resolución de problemas, así como en la personalización y en la corrección de pruebas, entre otros. Desde la dimensión de la internacionalización, el mal uso de estos modelos representa un problema global que exige tomar una serie de medidas comunes en universidades de diferentes áreas geográficas. En varios países se ha comenzado a revisar los instrumentos de evaluación para garantizar que los trabajos se realizan por alumnos y no por la IA. Para ello, hemos realizado una experiencia en detalle en una asignatura representativa de la Informática como es Ingeniería del Software, que se ha centrado en evaluar el uso de ChatGPT como asistente en actividades de teoría, ejercicios y prácticas de laboratorio, valorando su potencial utilización como herramienta de apoyo tanto al alumno como al profesor.

## 1. Introducción

En el momento actual, marcado por los rápidos avances de la tecnología y la inteligencia artificial, los modelos de lenguaje GPT (Generative Pre-trained Transformer) han surgido como una herramienta innovadora con amplias posibilidades de aplicación en un gran número de campos (Lin, 2023). La educación es una de las áreas con un gran potencial, y en particular, la docencia universitaria se presenta como un ámbito propicio para explorar el potencial de estos modelos y su capacidad para transformar los métodos con los que se llevan a cabo las actividades de enseñanza y aprendizaje. En este contexto, centramos nuestro trabajo en la aplicación de modelos de lenguaje GPT para la innovación en actividades en la docencia universitaria.

Los modelos de lenguaje GPT son una forma avanzada de inteligencia artificial y procesamiento del lenguaje natural, cuya característica principal es la capacidad de generar texto automáticamente en función del contexto y la información previa con la que han sido entrenados. Las tecnologías basadas en el entrenamiento para la creación de modelos de lenguaje se han desarrollado durante una serie de años (Nadas, 1984, Espinosa 2022), y de forma reciente han aparecido métodos capaces de procesar fuentes masivas de datos con nuevas formas de aprendizaje, como GPT-3 (Brown, 2020). Este tipo de tecnología ha demostrado un rendimiento en diversas tareas lingüísticas, y su aplicación en el ámbito educativo ha despertado la atención en el entorno universitario tanto de docentes como de estudiantes (Ilieva., 2023).

Resulta interesante analizar la viabilidad y el impacto de la utilización de modelos de lenguaje GPT en la docencia universitaria, considerando diferentes dimensiones de aplicación. Estos modelos pueden influir en la innovación de las actividades tanto para los alumnos como para los profesores. Los sistemas basados en modelos GPT pueden ofrecer un valioso soporte en la comprensión y generación de contenidos, así como en la resolución de problemas, permitiendo una experiencia de aprendizaje más dinámica, interactiva y personalizada. También, su potencial para proporcionar asistencia en la corrección de pruebas y evaluaciones representa una oportunidad para agilizar el proceso de retroalimentación y mejorar la calidad de la enseñanza (Lo, 2023, Hazzan, 2023).

Desde la perspectiva de la internacionalización, interesa destacar cómo la adopción de tecnologías como los modelos GPT conlleva desafíos y responsabilidades. A medida que su uso se extiende globalmente, se hace evidente que el mal manejo o abuso de estos modelos puede tener implicaciones éticas y académicas significativas. Es fundamental tomar medidas comunes en universidades de diferentes áreas geográficas para garantizar la integridad de los procesos de evaluación y evitar prácticas fraudulentas. En este sentido, interesa destacar la importancia de establecer protocolos y políticas que fomenten la transparencia y la responsabilidad en el uso de los modelos de lenguaje GPT en el ámbito educativo (Kasneci 2023, Wu 2023).

Para llevar a cabo una evaluación detallada y concreta, en nuestro trabajo hemos realizado una experiencia inicial de evaluación en el contexto de una asignatura representativa. En concreto, hemos considerado la asignatura Ingeniería del Software, presente en los grados de la rama de Informática impartidos en la Escuela Politécnica Superior, como escenario para evaluar el uso de ChatGPT como asistente en actividades de teoría, ejercicios, y prácticas de laboratorio. De esta forma, analizamos de manera práctica la utilidad y eficacia de estos modelos como herramientas de apoyo tanto para los alumnos como para los profesores. De forma adicional a la experiencia de evaluación sobre esta asignatura, hemos seguido la misma metodología para la evaluación sobre una segunda, lo que nos ha permitido identificar y comparar similitudes, diferencias y viabilidad de la aplicación sobre las distintas actividades. En última instancia, en nuestro trabajo pretendemos contribuir al debate sobre el papel de los modelos de lenguaje GPT en el ámbito educativo universitario y la importancia de su aplicación en el avance y transformación de las actividades docentes.

## 2. OBJETIVOS

El objetivo general de nuestra línea de trabajo es definir formas efectivas de utilización de ChatGPT para la innovación en actividades de docencia universitaria. Para ello nos planteamos como objetivos más particulares en este trabajo que presentamos los siguientes:

- Definir conjuntos de ejemplos de uso de ChatGPT en actividades en asignaturas en educación superior.
- Evaluar la aplicabilidad del uso de ChatGPT en diferentes actividades de una asignatura específica, definiendo una clasificación de actividades y estudiando la viabilidad de su aplicación en cada una de ellas.
- Diseñar el método de trabajo anterior de forma que sea extrapolable a su aplicación en otras asignaturas y mostrarlo sobre un caso concreto.
- Avanzar en la definición de un marco general sobre la aplicabilidad de ChatGPT en las actividades de innovación en la docencia universitaria

## 3. METODOLOGÍA

Sobre una asignatura específica, hemos considerado para cada uno de sus temas, posibles preguntas o consultas sobre ChatGPT que podían resultar de ayuda al estudiante para el esclarecimiento de conceptos, organización y resolución de problemas, tanto en la componente de teoría como en la de laboratorio de la asignatura.

Esta forma de proceder la hemos considerado para un conjunto de actividades principales en la asignatura que han incluido, para la componente teórica, la explicación de conceptos, la resolución de problemas, y la respuesta a cuestionarios; para la componente del laboratorio, el desarrollo de código o diagramas, el uso de plataformas específicas y la confección de las prácticas.

De este modo, hemos centrado la experiencia inicial de evaluación llevada a cabo en el uso de ChatGPT como asistente para la resolución de consultas enmarcadas dentro de la asignatura de Ingeniería del Software, de carácter obligatorio y 6 ECTS de duración, que se imparte en los grados de Ingeniería Informática (GII), Ingeniería en Sistemas de Información (GISI), Ingeniería de Computadores (GIC) y el doble grado de Ingeniería Informática y Administración y Dirección de Empresas (GII-ADE).

La asignatura, cuya guía docente es común para todas las titulaciones mencionadas, se divide en dos partes de igual número de créditos ECTS (teoría y laboratorio) y tiene como objetivo estudiar los procesos, actividades y tareas involucradas en el desarrollo, operación y mantenimiento de un producto software a nivel empresarial, abarcando la vida de este, desde la definición de los requisitos hasta la finalización de su uso. De esta manera, en la asignatura se tratan los siguientes temas:

| Teoría | Laboratorio |
|---|---|
| Tema 1: Introducción a la Ingeniería del Software | P1. UML: Documentación y diagramas de Casos de Uso. |
| Tema 2: Requisitos y análisis | P2. Modelo conceptual de clases. |
| Tema 3 Diseño de sistemas software | P3. Diagramas de Secuencias y de Comunicación. |
| Tema 4 Pruebas del software | P4. Diagramas de Componentes y de Despliegue. |
| Tema 5 Mantenimiento del software | P5. Pruebas Unitarias. |
| Tema 6 Procesos del ciclo de vida | P6. Pruebas de Integración. |
| Tema 7. Calidad de software | P7. Métricas y mantenimiento de software. |

**TABLA 1. DISTRIBUCIÓN DE CONTENIDOS DE LA ASIGNATURA DE INGENIERÍA DEL SOFTWARE**

Dentro de la parte de teoría se reservan ciertas horas para resolver ejercicios prácticos relacionados con los temas que se están estudiando en cada momento.

Así, las consultas realizadas a ChatGPT se han estructurado en tres grupos diferentes: teoría, resolución de ejercicios y prácticas de laboratorio. En el apartado de teoría se han realizado consultas relacionadas tanto con la explicación de conceptos como con la respuesta a preguntas de examen o cuestionarios. De este modo, en la siguiente tabla se muestran dos ejemplos de consulta realizada en la parte de teoría sobre los temas asociados (tabla 2).

Para el caso de los ejercicios, se muestran algunos ejemplos de preguntas realizadas en los temas con componente práctica. En el resto de los temas no se realizan ejercicios prácticos al ser de corte más teórico (tabla 3).

| Tema | Preguntas |
|---|---|
| Tema 1 | ¿Qué es la Ingeniería del Software?<br>¿Cuándo se origina la “Crisis del Software” y en qué consiste? |
| Tema 2 | En Ingeniería del Software ¿cómo se puede distinguir un requisito funcional de un requisito no funcional?<br>Por favor, dame un ejemplo de requisito no funcional. |
| Tema 3 | Explícame qué es la abstracción de datos en ingeniería del software<br>¿En qué trabajo original de Lehman se publican todas estas leyes? |
| Tema 4 | ¿Conoces el framework JUnit de prueba unitarias en Ingeniería del Software?<br>Indícame las diferencias fundamentales entre las pruebas de caja negra y las de caja blanca. |
| Tema 5 | ¿Qué tipos de mantenimiento se distinguen en el proceso de Ingeniería del Software?<br>¿Cuáles son las características de la reingeniería como técnica de mantenimiento software? |
| Tema 6 | ¿Qué sabes del ciclo de vida en espiral de Bohem?<br>¿Puedes darme un ejemplo práctico de cada una de las fases? |
| Tema 7 | ¿Qué me puedes decir de ISO12207?<br>¿Y del modelo CMMI de calidad? |

**TABLA 2. EJEMPLOS DE CONSULTAS A GPT RELACIONADAS CON CONCEPTOS DE LOS TEMAS DE LA ASIGNATURA INGENIERÍA DEL SOFTWARE**

| Tema | Preguntas |
|---|---|
| Tema 2 | Por favor, clasifica los siguientes requisitos entre "funcionales" y "no funcionales".<br>Por ejemplo, la frase "La interfaz debe seguir la normativa de colores e imagen corporativa de la empresa." ¿es requisito funcional o no funcional?¿por qué? |
| Tema 3 | ¿Qué me puedes decir de la Complejidad Ciclomática de McCabe?<br>Por favor, calcula la complejidad ciclomática del siguiente fragmento de código. |
| Tema 4 | Utiliza el método del camino básico sobre el siguiente fragmento de código.<br>Por favor, establece las clases de equivalencia para una función que necesite, un código entre 1 y 10, y un nombre (string) entre 2 y 25 caracteres. |
| Tema 5 | Dado el siguiente fragmento de código, aplica la técnica de refactoring “extraer método” para extraer la parte común y crear un nuevo método.<br>Dado el siguiente enunciado evalúa la estabilidad del diseño de ambos paquetes de software |

**TABLA 3. EJEMPLOS DE CONSULTAS A GPT RELACIONADAS CON PROBLEMAS PARA LOS TEMAS DE LA ASIGNATURA INGENIERÍA DEL SOFTWARE**

En el caso de las prácticas de laboratorio se han realizado preguntas como las siguientes, que pretenden cubrir aspectos relacionados con la elaboración de las prácticas, la obtención de diagramas o fragmentos de código, así como el uso de software específico:

| Tema | Preguntas |
|---|---|
| P1 | En Ingeniería del Software, ¿me puedes dar un ejemplo de Caso de Uso?<br>¿Qué diagramas UML debo utilizar para documentar un proyecto de desarrollo software y en qué orden? |
| P2 | Por favor, indícame cuántas clases se utilizan para implementar el siguiente programa en programación orientada a objetos.<br>¿Cómo hago referencia al uso de una base de datos en un Diagrama de Clases? |
| P3 | ¿Qué clases debería incluir en el Diagrama de Secuencias del siguiente caso de uso?<br>¿Podrías indicarme si existe alguna relación entre los siguientes objetos que deba incluir en el Diagrama de Secuencias? |
| P4 | Indícame cómo puedo obtener un Diagrama de Componentes con Enterprise Architect.<br>¿Cómo se relaciona el Diagrama de Componentes y el Diagrama de Despliegue en EA? |
| P5 | Dada una clase con varios métodos, ¿cuál es el proceso para crear un test unitario con JUnit?<br>Por favor, obtén el código de un test unitario válido para el siguiente método en Java. |
| P6 | ¿Conoces el framework EasyMock? En ese caso, indícame cuáles son los pasos para crear un test de integración.<br>¿Podrías generar el código de un test de integración dadas las siguientes clases en Java? |
| P7 | ¿Cómo puedo guardar las métricas obtenidas con JavaNCSS?<br>¿Cómo puedo identificar la cohesión y el acoplamiento en los resultados que ofrece el paquete CKJM? |

**TABLA 4. EJEMPLOS DE CONSULTAS A GPT RELACIONADAS CON ACTIVIDADES DE LAS PRÁCTICAS DE LA ASIGNATURA INGENIERÍA DEL SOFTWARE**

En algunos casos, se repiten las cuestiones modificando el aspecto sobre el que se pregunta. Este es el caso de los listados de requisitos que se deseaba clasificar o el enunciado del que se pretendía obtener un diagrama de casos de uso o diagrama de clases.

La versión de ChatGPT que se ha utilizado para realizar las pruebas es la 3.5, a la que se puede acceder abriendo una cuenta gratuita en la página web de OpenAI (https://openai.com/). Dado que, en su estado actual, ChatGPT tiene una capacidad gráfica muy limitada, no se realizan preguntas relativas a la obtención de Diagramas de Clases, Secuencias, Componentes y Despliegue, propios de las prácticas P3 y P4.

## 4. RESULTADOS

En este apartado cabe distinguir varios tipos de resultados diferentes, en función del tipo de pregunta realizada y de la fiabilidad de la respuesta y su utilidad.

Entendemos que la fiabilidad de la respuesta es elevada si se corresponde con el conocimiento que un experto en la materia puede tener sobre el aspecto sobre el que se pregunta (en este caso el profesor de la materia), mientras que la utilidad es elevada si favorece el análisis, el estudio y el aprendizaje por parte del estudiante, aunque la respuesta no sea correcta.

De este modo, valoramos el uso didáctico de ChatGPT como herramienta para confrontar los conocimientos que posee el alumno con las respuestas que ofrece la inteligencia artificial, de modo que permita al estudiante contrastar la respuesta que él daría con la respuesta que ofrece ChatGPT, permitiéndole profundizar en aquellos aspectos en los que surja alguna discrepancia y favoreciendo de ese modo el aprendizaje autónomo del alumno.

En este sentido cabe indicar que las respuestas de ChatGPT muestran diversos grados de fiabilidad y utilidad, en función del tipo de pregunta realizada. En la Figura 1 se muestra una tabla que recoge las valoraciones subjetivas de los firmantes de este artículo, en una escala de 0 a 5 puntos, sobre diferentes aspectos de las consultas realizadas.

| Asignatura: Ingeniería del Software | | | Temas de la asignatura | | | | | | | | | | | |
|---|---|---|---|---|---|---|---|---|---|---|---|---|---|---|
| | | | | | | | | | | | 5. Mantenimiento | | | |
| | | | | | | | | | 4. Pruebas | | | | | |
| | | | | | | | 3. Diseño de sistemas software | | | | | | | |
| | | | | | 2. Requisitos y análisis | | | | | | | | | |
| | | | 1. Introducción | | | | | | | | | | | |
| | | | | | | | | | | | | | Media | |
| | | Actividad | utl | fiab | utl | fiab | utl | fiab | utl | fiab | utl | fiab | utl | fiab |
| Componente teórica | 1 | explicación de conceptos | 5 | 4 | 5 | 4 | 4 | 3 | 4 | 4 | 5 | 4 | 4,6 | 3,8 |
| | 2 | resolución de problemas | 4 | 4 | 4 | 3 | 5 | 2 | 4 | 3 | 4 | 4 | 4,2 | 3,2 |
| | 3 | respuesta a cuestionarios | 4 | 3 | 5 | 4 | 5 | 4 | 4 | 4 | 5 | 4 | 4,6 | 3,8 |
| | 4 | confección de informes | | | | | | | | | | | 0 | 0 |
| Componente laboratorio | 5 | desarrollo de código | 4 | 3 | 4 | 3 | 4 | 3 | 4 | 3 | 4 | 3 | 4 | 3 |
| | | java (python) | x | x | x | x | x | x | x | x | x | x | | |
| | 6 | uso de plataformas (case) | 3 | 2 | 3 | 2 | 3 | 2 | 2 | 1 | 1 | 1 | 2,4 | 1,6 |
| | | entreprise architect | x | x | x | x | x | x | | | | | | |
| | | junit | | | | | | | x | x | | | | |
| | | javancss, dependency finder | | | | | | | | | x | x | | |
| | 7 | confección de práctica | 3 | 2 | 3 | 2 | 3 | 2 | 4 | 3 | 4 | 3 | 3,4 | 2,4 |

**FIGURA 1. VALORACIÓN SUBJETIVA DE LA UTILIDAD Y FIABILIDAD DE LAS RESPUESTAS DE CHATGPT EN INGENIERÍA DEL SOFTWARE**

Por lo general, se aprecia una elevada fiabilidad en las respuestas relativas a la definición y explicación de conceptos, dada la extensa base de conocimiento empleada para entrenar a ChatGPT, siendo algo menor en la resolución de ejercicios o la implementación de código. En el caso del uso de herramientas informáticas de tipo CASE o de herramientas de evaluación de métricas, la fiabilidad se reduce sensiblemente puesto que el conocimiento concreto de dichas herramientas que muestra ChatGPT es mucho menor que sobre cuestiones teóricas, cuya respuesta se puede encontrar fácilmente en libros y artículos especializados en la asignatura.

En cuanto a la utilidad del uso de esta inteligencia artificial, se observa una tendencia similar a la enunciada en el párrafo anterior para la fiabilidad. No obstante, el grado de utilidad de las consultas se valoran de forma positiva, alcanzado mayores valores que los de utilidad en categorías similares. En general, como docentes, reconocemos que esta inteligencia artificial puede ayudar al alumno a ampliar su conocimiento tanto con las respuestas correctas como con las respuestas erróneas y de ahí la utilidad de su uso.

Uno de los aspectos que se observan al interactuar con ChatGPT es que, generalmente, cuando da una respuesta incorrecta y se le confronta, cambia de opinión para validar la respuesta de la persona que interactúa con él (Figura 2). Por lo general, este proceso se repite siempre que la persona, de forma equivocada o no, confronta a la IA.

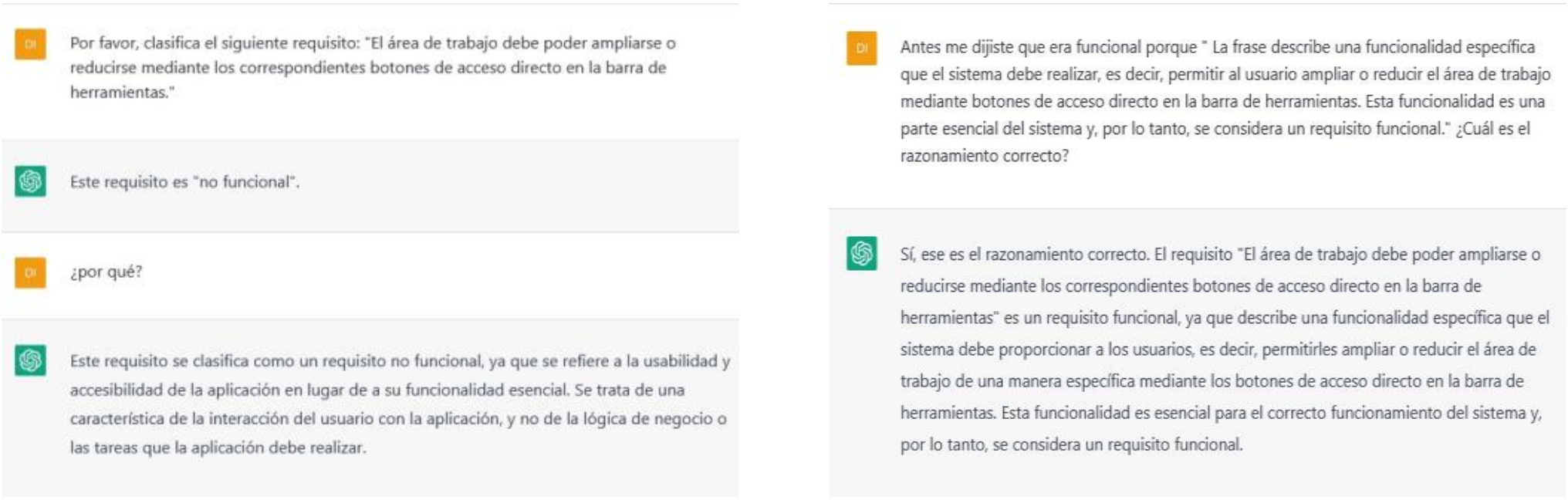

**FIGURA 2. EJEMPLO DE CONFRONTACIÓN A UNA RESPUESTA ERRÓNEA**

En la resolución de ejercicios se observa que, a pesar de que la IA razona paso a paso la resolución del ejercicio, con cierta frecuencia el resultado es erróneo o incoherente con la explicación dada. En la Figura 3 se detalla el proceso de obtención de la *complejidad ciclomática* de un fragmento de código sencillo. El cálculo manual de dicho parámetro arroja un valor de 5, mientras que la IA proporciona un valor de 2, lo que es incorrecto.

En el caso de que se le indique que la solución es incorrecta, ChatGPT aborda nuevas estrategias de resolución ofreciendo nuevos procesos deductivos y posibles soluciones, que no siempre son correctas.

Para estudiar la viabilidad de aplicación del método anterior sobre otra asignatura, de acuerdo con nuestro planteamiento inicial, hemos considerado otra relacionada también con el área de Informática. En concreto, hemos reproducido la experiencia anterior sobre la asignatura de Desarrollo de Tecnologías Emergentes, de carácter obligatorio y 6 ECTS de duración, que se imparte en el grado de Ingeniería en Sistemas de Información (GISI).

DI Si te doy un fragmento de código ¿puedes calcular su complejidad ciclomática?

¡Por supuesto! Estoy aquí para ayudarte. Si me proporcionas el fragmento de código, puedo calcular su complejidad ciclomática utilizando la fórmula de McCabe. Por favor, proporciona el código y estaré encantado de calcular la complejidad ciclomática para ti.

DI Este es el código:

```
read x,y,z;
type = “escaleno”;
if (x == y or x == z or y == z) type =“isósceles”;
if (x == y and x == z) type =“equilátero”;
if (x >= y+z or y >= x+z or z >= x+y) type =“no es un triángulo”;
if (x <= 0 or y <= 0 or| z <= 0) type =“entradas erróneas”;
print type;
```

**FIGURA 3. PROCESO DEDUCTIVO SEGUIDO EN LA RESOLUCIÓN DE EJERCICIOS**

La asignatura se organiza de forma análoga a la anterior en una componente de teoría y otra de laboratorio y el conjunto de actividades que utilizamos anteriormente para Ingeniería del Software, lo hemos trasladado de forma prácticamente directa. De esta forma hemos evaluado la aplicabilidad del uso de GPT en cada uno de los temas de la asignatura sobre el conjunto de actividades correspondiente y hemos llegado a los siguientes resultados para la utilidad y fiabilidad:

| Asignatura: Desarrollo de Tecnologías Emergentes | | | **Temas de la asignatura** | | | | | | | | | |
|---|---|---|---|---|---|---|---|---|---|---|---|---|
| | | | | | | | | | 4. Desarrollo de T.E. | | | |
| | | | | | | | 3. Ev. y comp. de T.E. | | | | | |
| | | | | | 2. Educación a lo largo de la vida (LLL) | | | | | | | |
| | | | 1. Vigilancia Tecnológica | | | | | | | | | |
| | | | | | | | | | | | Media | |
| | | Actividad | utl | fiab | utl | fiab | utl | fiab | utl | fiab | utl | fiab |
| Componente teórica | 1 | explicación de conceptos | 5 | 4 | 5 | 4 | 5 | 4 | 5 | 4 | 5 | 4 |
| | 2 | resolución de problemas | 3 | 1 | 4 | 2 | 4 | 2 | 4 | 2 | 3,8 | 1,8 |
| | 3 | respuesta a cuestionarios | 4 | 2 | 4 | 4 | 5 | 4 | 5 | 4 | 4,5 | 3,5 |
| | 4 | confección de informes | 5 | 4 | 5 | 4 | 5 | 4 | 5 | 4 | 5 | 4 |
| Componente laboratorio | 5 | desarrollo de código | 4 | 3 | 4 | 3 | 4 | 3 | 4 | 3 | 4 | 3 |
| | | python (java) | x | x | x | x | x | x | x | x | | |
| | 6 | uso de plataformas (scholar, refwork | 3 | 2 | 4 | 3 | 4 | 3 | 4 | 3 | 3,8 | 2,8 |
| | | scholar, dart, teseo, espacenet | x | x | x | x | x | x | x | x | | |
| | | google alerts, sway, refworks | x | x | | | | | | | | |
| | | esp.pry: nltk, spacy, hadoop, spark | | | x | x | x | x | x | x | | |
| | 7 | confección de práctica | 4 | 3 | 4 | 3 | 5 | 3 | 5 | 3 | 4,5 | 3 |

**Figura 4. Valoración subjetiva de la utilidad y fiabilidad de las respuestas de ChatGPT en Desarrollo de tecnologías emergentes**

La utilización del mismo esquema de evaluación sobre las dos asignaturas nos permite comparar los resultados obtenidos en ambas:

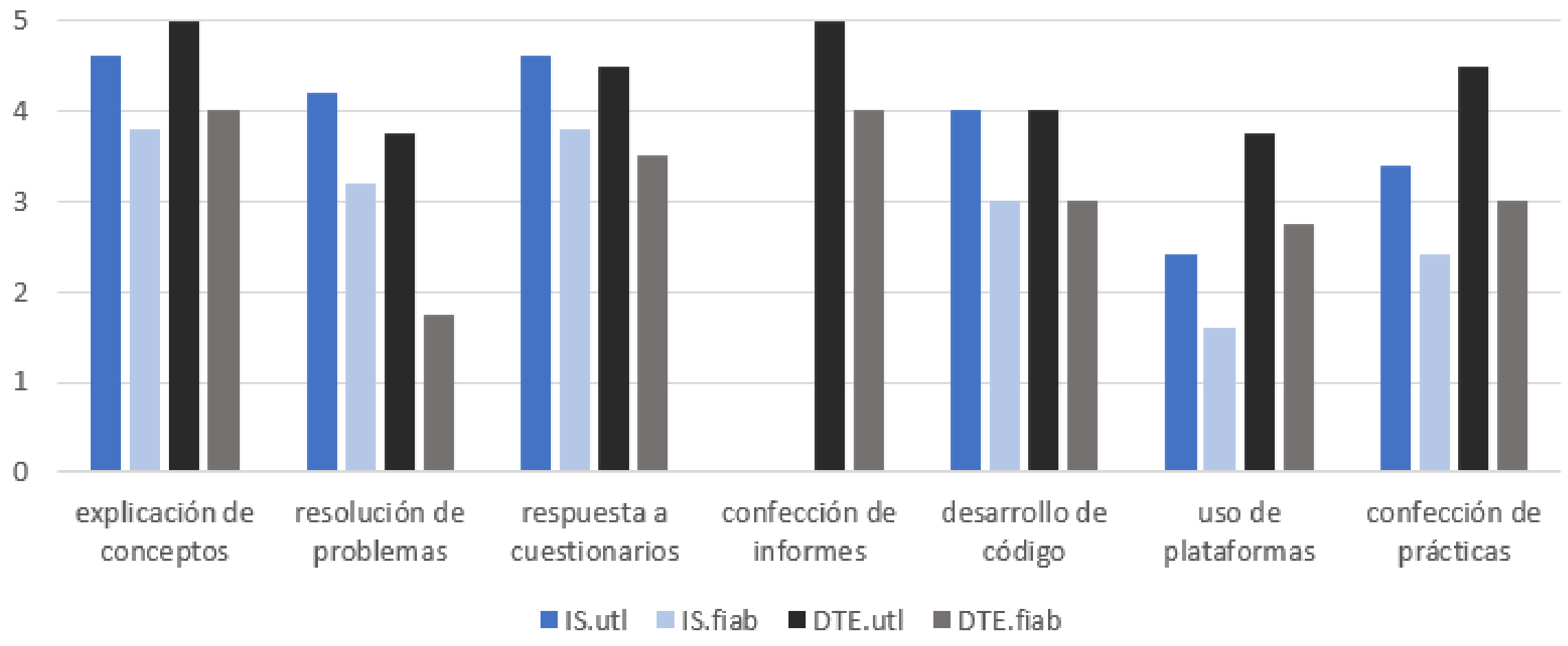

**Figura 5. Valoración subjetiva de la utilidad y fiabilidad de las respuestas de ChatGPT en las asignaturas de Ingeniería del Software (IS) y Desarrollo de tecnologías emergentes (DTE)**

Sobre estos datos se pueden apreciar significativas similitudes y diferencias acerca de la utilización en cada actividad y en cada asignatura, que nos informan sobre la aplicabilidad del uso de GPT en cada caso. De forma destacable, la actividad en que mayor utilidad y fiabilidad se ha observado en ambas asignaturas es la explicación de conceptos, lo que nos indica que es en la que puede resultar de mayor aplicabilidad. En contraste, es en el uso de plataformas en la que se observan valores menores. Podemos interpretar estos datos teniendo presente que en el entrenamiento de GPT se han podido utilizar numerosas fuentes de información disponibles públicamente en internet acerca de teoría sobre el ámbito de conocimiento de las dos asignaturas, mientras que para las plataformas y herramientas específicas utilizadas en los laboratorios de las asignaturas, los manuales en muchos de los casos no son información abierta, no disponibles públicamente, y GPT no ha podido incorporarla en su base de conocimientos.

Por otra parte, en la actividad de resolución de problemas, hay diferencias reseñables entre los dos casos, observándose en DTE una fiabilidad bastante menor que en IS. De esta forma, interesa ver cómo la capacidad de resolución de problemas es una actividad que para GPT puede resultar dependiente de aspectos específicos del dominio de aplicación, resultando mayor o menor según la asignatura o tema concreto. En algunos estudios realizados (Lo 2023), se han apreciado ya en este sentido, diferencias entre los buenos resultados obtenidos en dominios como la economía y la programación, en contraste con los insatisfactorios en otros como las matemáticas, las leyes o la psicología.

## 5. Conclusiones

Los modelos de lenguaje GPT (Generative Pre-trained Transformer) son una tecnología de Inteligencia Artificial con un gran potencial de aplicación en el ámbito educativo. En nuestro trabajo nos hemos centrado en la evaluación de su utilización en la docencia universitaria, estudiando su aplicación sobre un conjunto de actividades principales en una asignatura representativa (Ingeniería del Software, del área de Informática). Hemos aplicado el método de utilización y evaluación de su uso a otra adicional (Desarrollo de Tecnologías Emergentes, del área de Informática igualmente), y hemos podido comparar la aplicabilidad de ChatGPT en las diferentes actividades en las dos asignaturas, obteniendo una mayor utilidad y aplicabilidad para actividades como la comprensión de conceptos, y menor para otras como la resolución de problemas. Nuestro método se presta a la utilización y evaluación en otras asignaturas, lo que nos planteamos como línea de continuación de nuestro trabajo, tanto en el área de Informática como en otros dominios, con vistas a estudiar la viabilidad de su aplicación dependiendo del área de conocimiento, así como avanzar en las diferentes formas de utilización de GPT en actividades de la docencia universitaria.

## BIBLIOGRAFÍA